# Research on Potential Semantic Web Service Discovery Mechanisms

Anji Reddy A[1] and Sowmya Kamath[2]
*Department of Information Technology,*
*National Institute of Technology Surathkal, Mangalore, India*
anjireddy6@gmail.com[1]    sowmyakamath@ieee.org[2]

**Abstract:**
Web services are an important paradigm in distributed application development. Currently, many businesses are seeking to convert their applications into web services because of its ability to promote interoperability among applications. As a number of web services increase, the process of discovering appropriate web services for consumption from user's perspective gains importance. In this paper, we present a study of prospective ways of discovering web services and issues related to each of them. In addition, we discuss ontology concepts and related technologies, which incorporate semantic meaning and hence give domain knowledge about a web service to improve the discovery mechanism. The paper also presents an overview of related research work, identifying metrics useful in filtering web service search mechanisms.

## 1. INTRODUCTION

During the past decade, a lot of researchers have directed significant interest towards web services, an important paradigm of Service–Oriented Architecture (SOA). SOA is an emerging technology in the development of loosely coupled distributed applications on the web. Web services are one of the techniques to implement SOA, a software system designed to support interoperable machine-to-machine interaction over a network [1] (i.e. Browser and Platform independent). Web services convert software applications into web applications, provide loose-coupling at middle-ware level and open an interface to consumers without making them aware of the underlying technologies or implementation details.

The complexity of web services vary in function from simple applications such as weather reports, currency convertors, credit checking, credit card payment, etc to complex business applications like those of Online Book stores, insurance brokering system, online travel planners etc. Current availability of web services in repositories such as UDDIs, Web portals (e.g. Xmethods [2], webservicex [3], webservicelist [4] etc), over the Internet and a further scope of rapid future increases makes discovery of web services an important issue from user's perspective.

The discovery of potential web services can be possible mainly through two approaches; one is by using centralized repositories such as UDDI's and another one is by using web crawling techniques. UDDI constitutes metadata about web services and fulfils advertisement requirements to service providers. It also provides search facilities to users for publishing and invoking services. Various standards such as WSDL, SOAP, and UDDI have been developed to support discovery of web services, but pure syntactic nature of all these standards yields in-efficient search mechanism. In addition of semantics to the web services through ontology related technologies can improve the above mentioned syntactic related search.

The rest of the paper is organised as follows: In section 2, we present a brief overview of web service architecture and its protocol stack for easy flow of communication and connection among applications. Section 3 highlights the potential mechanisms for web service discovery. In Section 4, we discuss the UDDI searching mechanism highlighting the existing issues, and also addition of semantics through ontology related technologies. In Section 5, we present web service crawlers as an alternative technique for web service discovery and present a few search filtering mechanisms in section 6, followed by conclusion and references.

## 2. PRELIMINARY CONCEPTS OF WEB SERVICE ENGINEERING

The following Web Service Architecture (Fig 1) and Protocol Stack define the roles played by different entities in web service transaction and technologies.

### 2.1 Web Service Roles:

Web service plays mainly three different kinds of roles in the architecture as follows [5]:
- *Service Registry*- It is a logical centralized directory structure for services, provides a central place where service providers can advertise (Publish) their new services and/or discovers existing services.
- *Service Provider* - It defines a service description for the web service and publishes it into a service registry.
- *Service Consumer*- The consumer of a web service looks up the service registry for a particular service and retrieves the service description to bind with the web service provider and invokes/ interacts with the service implementation.

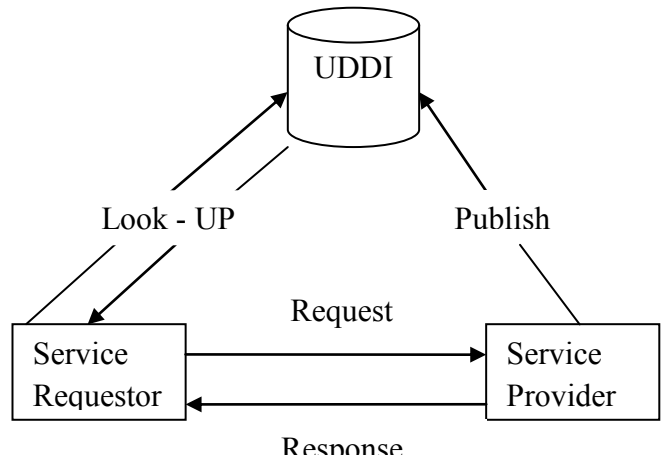

**Fig 1. Overview of Web Service Architecture.**

### 2.2 Web Service Protocol Stack:

The protocol stack defines the process of web service involves in network communications by using network technologies such as HTTP, HTTPS transport protocols, the SOAP message protocol, the XML language, UDDI and WSDL to complete the connection between applications [6][7].SOAP is a light weight protocol which is responsible for the exchange of information over a distributed network independent of platform and language. SOAP uses HTTP and XML as the mechanisms for information exchange and is a format for sending messages and makes resource or object access by allowing applications to invoke object methods or functions, residing on remote servers. It consists of an envelope that defines a framework for describing what is in a message and how to process it, encoding rules and convention for representing remote procedure calls/functions.

WSDL is a simple XML based language used to describe the web services and gives information about how to access them in decentralized and distributed environments. The XML format of WSDL document describes the web service based on major elements such as Types, Messages, Port Type, Binding and Service [8]. The Types element contains the XML schema data type definitions used by the web service and the Message element defines input/output parameters of a web service operation, moreover each Message can consist of one or more parts. A part is like parameters of a function call in traditional programming languages and each part has a unique name and type definitions. Port Type describes a group of operations performed by the web service and the messages that are involved. Port Type is just like a function library (or a module or a class) in traditional programming languages [9]. The Binding element describes the details of each and every port type message format and its protocol and Service element defines a collection of ports or end points that describes a specific binding.

UDDI is a platform independent centralized distributed XML based business registry used to publish and discover web services. It's an open industry standard initiated by OASIS in 2000. The UDDI business registry is intended to serve as a global, all-inclusive listing of businesses and their services and it doesn't contain detailed specification about business service but it directs to resources that contain the actual service implementation. Core information model used by UDDI registries is defined in an XML schema and offers businesses to advertise in three kinds of components called White pages, Yellow pages and Green pages [10].

## 2.3 Potential Discovery Mechanisms

Currently, approximately 63% of available web services on the web are considered to be active [11]. So, discovery of relevant and useful web services is an important aspect to provide better quality of service to consumers. Currently there are two approaches exist for web service discovery -
1. Web Service Discovery through Universal Description, Discovery and Integration (UDDI)
2. Web Service Discovery Based on Web Crawlers.

# 3. WEB SERVICE DISCOVERY THROUGH UNIVERSAL DESCRIPTION, DISCOVERY AND INTEGRATION

UDDI registries were introduced in the year August 2000 by IBM, Microsoft and SAP. It bridges the gap between service providers and consumers, helps in discovery and invoking of services through a public or private dynamic brokerage system [10][12]. Each business registered with UDDI categorises all of its web services according to a defined list of service types called taxonomies. The UDDI search is based on metadata of services like service name, providers name and t-Models name.

Each service can have one or more t-Models that are used for describing the attributes and characteristics of a service. The UDDI t-Models represent a technical specification, typically a specification pointer or metadata about a specification document (WSDL File), which includes name and an URL pointing to an actual implementation of the WSDL document which describes the web service. Find and retrieval of web service information is possible through a group of SOAP messages represented by t-Model. The available UDDI inquiry API's functions are `find_binding`, `find_business`, `find_service`, `find_tModel, find_businessdetail` [13] [14] etc.

### 3.1. Issues Related to UDDI:

The UDDI registry has not been adopted in the way its designers hoped at the beginning. The providers of public UDDI's, IBM, Microsoft and SAP announced that they were closing their services in January 2006 [10][15][16]. Microsoft announced that they were removing UDDI services from the future version of Windows server operating system in September 2010 but continued the future releases of UDDI services as part of Biz Talk product [17]. Currently, there are no public UDDI's available, but any organization or person can use the private UDDI in a company intranet or within and between enterprises. UDDI searches web services based on key word matching, syntactically not on the semantics of the service it provides.

### 3.2. Semantic Web Service Engineering:

The lack of semantic focus in current web service technologies (WSDL, UDDI and SOAP) makes the searching mechanism more complex, which leads to an inefficient search. The Semantic web is an extension to the current classical "web of documents" which is understandable by humans to the "web of data", in which machines can communicate among themselves by providing semantic meaning to web content [18][19]. Semantic web services are similar to other web services in the client-server system; however they use the semantic meaning of functions it operates on. Web services are components or resources in the semantic web which helps machines read or understand the data in a detailed and sophisticated way.

According to W3C's vision, semantic web technologies enable users to create data stores on web, build vocabularies and write rules to process data. This linked data of web empowers technologies such as Ontology Interference Language (OIL), DARPA Agent Mark-up Language (DAML), DAML+OIL, Resource description Framework (RDF), Web Ontology Language (OWL), Web Services Modelling Language (WSML), Web Services Semantics (WSDL-S), SAWSDL, OWL-S. Ontology is the main building block for all the semantic web service technology languages.

### 3.3. Ontological Web Service Engineering:

The term "ontology" is used in many fields such as Artificial Intelligence, Software engineering, Information architecture and many more. Ontology is a structural framework for organizing information or knowledge representation about the world or part of a world, which describes a set of concepts, its properties and relationships among them. Ontology is a "formal specification of a shared conceptualization", provides a shared vocabulary, taxonomy of a particular domain which defines objects, classes, its attributes and their relationships.

Ontology provides semantics to web services and its resources and can greatly improve search mechanisms. In order to realize the vision of semantic web, researchers have proposed several languages, algorithms and architectures. Ontology based languages like OIL DAML, and OWL can be used to describe a web service and its resources. OWL-S and WSMO are standardized, richest and popular The OWL-S (OWL for Services), is an ontology for web services expressed in OWL and replaces the existing web services by adding semantics which is understandable by computer programs unlike a WSDL, which is a specification that supplies only what happens when web service is used [20][21]. OWL-S enables the automation of programs actively in service discovery, interoperation and service composition.

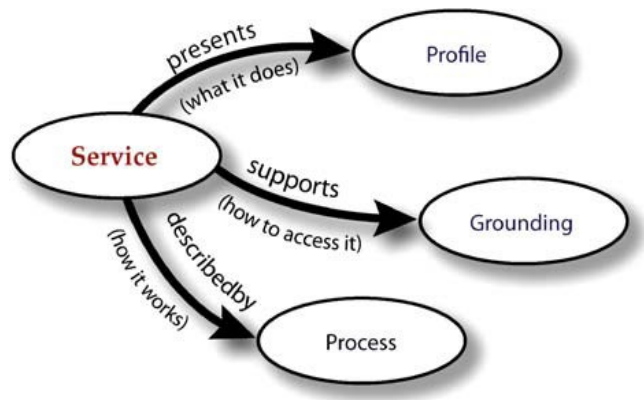

Figure 2. Service Ontology OWL-S [30].

The overall OWL-S structure explains about web service functions and the communication details. It contains three main parts called the service profile, the process model and the grounding [18]. The service profile used to describes about what the service does and gives abstract characterization information of a service such as service name, description, owner and contact details etc…, suitable for the purpose of advertisement and selection. At an abstract level, the service model provides a complete description of how to interact with service, information of inputs, outputs, preconditions and results. The service grounding provides information about *how it works,* it includes how to interact with service such as communication protocols, message formats, port no's etc. In addition to UDDI search mechanisms which are classification based ones; OWL-S provides capability-based search mechanisms. In capability-based search mechanisms, discovery of web services is based on the inputs and preconditions that needed to be satisfy and outputs and effects should be produced according to the user needs.

## 4. WEB SERVICE DISCOVERY BASED ON WEB CRAWLERS

An alternative mechanism to discovering of web services is through web crawling. The discovery mechanism of web service access points is no longer available through public service registries (i.e. UDDI) [16]. The common problems like centralized node failures and bottleneck issues in registries faced while using a centralized repository like the UDDI make for searching of web services through web crawling more attractive. Al-Masri and Mahmoud [22][23], state that web search engines such as Google can be effectively and efficiently used as an alternative to web service registries.

The web service crawler engines (WSCE), which crawls web to get the information from different accessible UBR's, service portals and search engines [24][25][26]. The web service portals or directories such as WebServiceList, RemoteMethods, WSIndex, and XMethods.net are the ones still active and providing services. However, these portals or directories do not implement original web service standards, like UDDI which doesn't give a guarantee of continuous and reliable services in future. Currently the public UDDI registries or directories maintained by IBM, Microsoft and SAP have been shut down, which only leaves the option of discovery of web services through private UBR's (maintained separately by organizations or individuals) and search engines[16][27].

The web service crawler engine (WSCE) procedure explains about the automatic discovery of a web service's WSDL or OWL-S documents which are handled just like a normal web page, which then collects web service information into a centralized repository called web service storage (WSS) by using search engine API's such as Google SOAP search API's. Further requests to web services can be obtained from the repository or cached WSS maintained earlier. Other approaches are crawling of web services based on service descriptions and using ontologies in the discovery process. However, WSCE discovery mechanism mainly follows crawling, processing, indexing and searching phases.

## 5. Web Service Filtering Mechanisms

Various studies in this area have focussed on how to improve the search results of web services and its accuracy by considering the goals of a user. Furthermore, the user requires web services that not only meet their required functionality but also the quality to satisfy their needs. So, web service quality attributes play a major role in querying the similar functionality web services in accomplishing the choosing of required web services [28]. Al Masri and Mahmoud [29] classify the quality of attributes into objective and subjective attributes.

The Objective parameters are corresponding to concrete or quantitative measurements, while the subjective parameters are measurements based on client's perception or regulated by the service provider. Performance measures are either based on response time or throughput and can narrow down the number of irrelevant web services. Dependability measures are either based on availability or reliability such as filtering out the broken link services or search relevant services. Conformance measures are based on best practices or compliance and Usability is based on documentation. The Subjective parameters are service value measurement based on price, compensation or penalty; and reputation, based on user ratings or feedbacks etc. Using these quality attributes in client queries helps in identifying their goals more accurately thereby significantly improving the service discovery process.

## 6. CONCLUSION

In this paper, we introduced the probable discovery mechanisms of web services, the web service architecture and its related technologies. The semantic web concept and its characteristics and the realization of semantics to web services through ontology and ontological languages were also discussed. We also presented a study of existing issues related to UDDI and how to overcome them by using the practical, reliable and an alternative approach to the centralized UDDI called web crawling for discovery of web services. Further we investigated the metrics used to filter web services in selection mechanisms.